\documentclass[useAMS,usenatbib]{mn2e}
\usepackage{amsmath}
\usepackage{placeins}
\usepackage{graphicx}
\usepackage{footnote}

\usepackage{natbib}

\usepackage{color}
\usepackage{colortbl}
\usepackage{times}
\begin{document}

\def\etal{{\it et al.\ }}
\def\teff{{T$_{\text{eff}}$\,\,}}
\def\lg{{$\log$(g)\,}}

\def\gtaprx{ \mathrel{ \vcenter{
      \offinterlineskip \hbox{$>$}
      \kern 0.3ex \hbox{$\sim$}    } } }

\def\ltaprx{ \mathrel{ \vcenter{
      \offinterlineskip \hbox{$<$}
      \kern 0.3ex \hbox{$\sim$}    } } }
      
\def\apj{{ApJ}}
\def\apjs{{ApJS}}
\def\apjl{{ApJL}}
\def\araa{{ARA\&A}}
\def\aap{{A\&A}}
\def\pasp{{PASP}}
\def\pasa{{PASA}}
\def\mnras{{MNRAS}}
\def\nat{{Nature}}
\def\pasj{{PASJ}}
\def\icarus{{Icarus}}
\def\memsai{Mem. Soc. Astron. Italiana}
\def\apss{{Ap\&SS}}

\def\etal{{\it et al.\ }}

\title[Implications of the Spectroscopic Abundances in Alpha Centauri A \& B]{Implications of the Spectroscopic Abundances in Alpha Centauri A \& B}

\author[Natalie R. Hinkel \& Stephen R. Kane]{Natalie R. Hinkel\thanks{email: natalie@caltech.edu} \& Stephen R. Kane\\
$^{}$NASA Exoplanet Science Institute, Caltech, MS 100-22, 770\\
  South Wilson Avenue, Pasadena, CA 91125, USA\\}

\pagerange{\pageref{firstpage}--\pageref{lastpage}} \pubyear{2012}

\maketitle

\label{first page}

\begin{abstract}
Regardless of their close proximity, abundance measurements for both stars in $\alpha$ Centauri by different groups have led to varying results.  We have chosen to combine the abundance ratios from five similar datasets in order to reduce systematic effects that may have caused inconsistencies.  With these collated relative abundance measurements,
we find that the $\alpha$ Cen system and the Sun were likely formed from the same material,
despite the [Fe/H] enrichment observed in the $\alpha$ Cen binaries: 0.28 and 0.31 dex, respectively. 
Both $\alpha$ Centauri A and B exhibit  
relative abundance ratios
that are generally solar, with the mean at 0.002 dex and 0.03 dex, respectively.  The refractory elements (condensation temperature $\gtaprx 900 K$) in each have a mean of -0.02 and 0.01 dex  and a 1$\sigma$ uncertainty of 0.09 and 0.11 dex, respectively.  Given the trends seen when analyzing the refractory abundances [X/Fe] with condensation temperature, we find it possible that $\alpha$ Centauri A may host a yet un-discovered planet.

\end{abstract}


\section{Introduction}
In order to understand the evolution of the solar neighborhood and the Milky way, we utilize the chemical compositions of stars.  
Thin-disk stars in the vicinity of one another are usually affected by the same astrophysical events, 
which are then recorded in the protoplanetary disk composition.  Through the analysis of their composition, mainly via theoretical models such as \citet{Woosley:1995p3481}, we are able to better constrain events that determined the initial mass, star formation rate, inherited composition, and stellar yields.

Despite a litany of work analyzing the stellar atmospheric parameters and metallicity of the $\alpha$ Centauri (Cen) visual binary system, there seems to be little consensus between the measurements.  Even questions regarding the similarity of the stars to the Sun, to each other, or with respect to certain elements are not consistent.  \citet{PortodeMello_2008}, the most recent of the authors to analyze the abundance ratios within this system, graphically showed a handful of datasets for $\alpha$ Cen A and the rather large abundance ratio variations between them (their Fig. 8 and references therein).

Because of the proximity of the system, we are able to compile literature abundance ratios determined for the two nearby stars, with respect to the Sun, similar to \citet{Ramirez_2010}.  This also allows us to analyze the formation of the binary system, as illuminated by the abundances.
The recent discovery of a terrestrial planet orbiting $\alpha$ Cen B \citep{Dumusque_2012} presents a unique case study for examining the elements found not only within one of the closest stars to the Sun, but also within a binary where one of the stars is an exoplanet host.  It is especially interesting because, to-date, there has been no confirmation of an exoplanet around $\alpha$ Cen A.  

\section{Reference Analysis}
Multiple literature sources have measured the spectroscopic abundance ratios for the $\alpha$ Cen system.  However, few of those authors have measured both the A and B stellar components.  After searching the literature (any exclusion was not intentional), we have found that only seven literature sources measured both stars for multiple elements: \citet{AllendePrieto:2004p476, Gilli:2006p2191, Laird:1985p1923, Neuforge_1997,PortodeMello_2008, Thevenin:1998p1499, Valenti:2005p1491}.  

If we are to analyze the relative abundances of these two stars, we must first understand the data sets before we combine them.  The abundance measurements taken by \citet{AllendePrieto:2004p476} were conducted using both the 2.7m telescope at McDonald Observatory and the ESO 1.52m dish on La Silla.  They determined the abundances of 16 elements within 118 stars via differential analysis.  The MARCS code \citep{Gustafsson:1975p4658} was utilized for modeling the stellar atmospheres.  While they did not investigate the effects of non-local thermodynamic equilibrium (NLTE), they did take into consideration hyperfine splitting for Cu, Sc II, Mn, Ba II, and Eu II.  Their derived effective temperature and specific gravity for both stars which are \teff = 5519, 4970 K and \lg = 4.26, 4.59, respectively.

\citet{Gilli:2006p2191} measured the abundances of 12 elements for 101 stars in the solar neighborhood.  Their spectra spanned 3800 \AA\ to \hbox{10,000\AA} across five different spectrographs, with considerable wavelength overlap in-between.  The standard LTE analysis was conducted for all elements via MOOG \citep{Sneden:1973p6104} and the ATLAS9 atmospheres
\citep{Kurucz:2005p4698}.  The effective temperatures, surface gravities,
microturbulence, and metallicity [Fe/H] were determined by \citet{Santos:2005p2866,Santos:2004p2996}.  For both stars, respectively, \teff = 5844, 5199 K and \lg = 4.30, 4.37.

\begin{figure*}
\centerline{\includegraphics[height=2.5in]{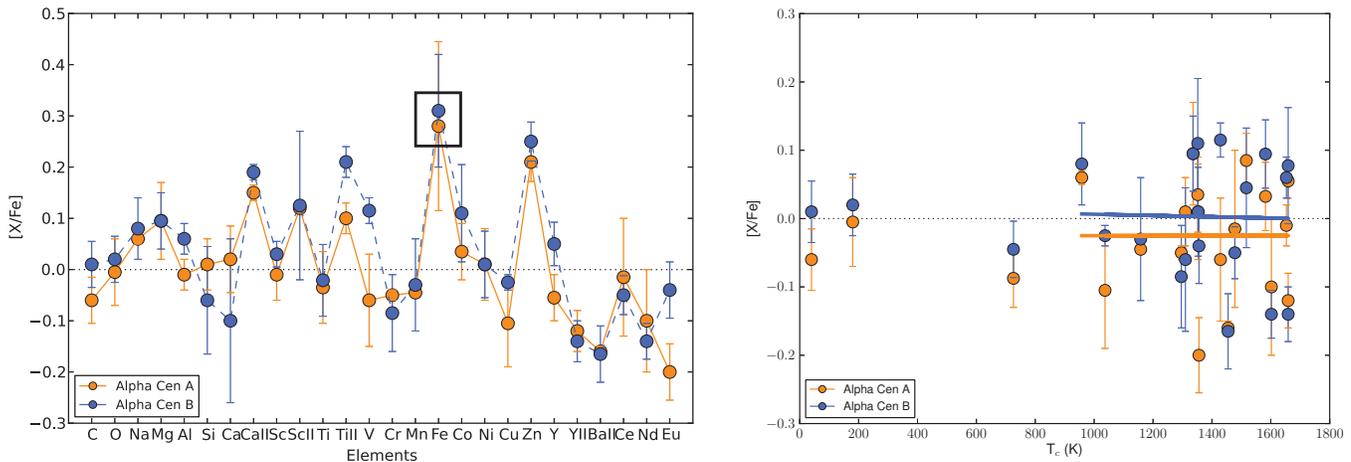}}
\caption{
Elemental abundance ratios for the A and B stars as rendered from five literature sources (left) and with respect to condensation temperatures (right). All elements are [X/Fe], with the exception of iron which is [Fe/H]
and denoted with a black box in order to avoid misinterpretation.
The dotted line denotes the solar values at 0.0 dex.  Errorbars depict the spread in the data when multiple catalogs measurements occurred for the element, standard uncertainty is given when only one catalog determined the abundance.  The solid lines (right) are a linear fit to the refractory elements ( T$_c$ $\gtaprx 900 K$), with slopes of 0.00015 $ \times 10^{-3}$ dex K$^{-1}$ and -0.0094 $ \times 10^{-3}$ dex K$^{-1}$, respectively.
}\label{fig.hist}
\end{figure*}

The abundances determined by \citet{PortodeMello_2008} for the $\alpha$ Cen system were extracted using differential analysis with respect to the Sun in order to reduce possibly NLTE effects. Their stellar atmospheres were determined via the NMARCS grid \citep{Edvardsson:1993p2124}, with discrepant results for \teff within the B-star between methods (see reference).  The found, respectively,  \teff = 5824, 5223 K and \lg = 4.34, 4.44.  Hyperfine corrections were included for Mg, Sc I, Sc II, V I, Mn, Co, Cu, and Ba II.

\citet{Thevenin:1998p1499} measured the abundances in 1108 late-type stars for 25 elements ranging from Li-Eu.  
Their analysis of these abundances is found in \citet{Thevenin1999}.  While they also examined the NLTE effects within predominantly metal-poor stars, they did not find any significant NLTE corrections for the abundances in solar-type stars, such as $\alpha$ Cen A and B.  The stellar parameters for both $\alpha$ Cen A and B, respectively, are \teff = 5727, 5250 K and \lg = 4.2, 4.6.

The work performed by \citet{Valenti:2005p1491} covered 1040 main-sequence stars for five elements, including iron.  
 They performed an SME analysis and used the ATLAS9 stellar model atmospheres \citep{Kurucz:2005p4698}, for which they determined the stellar parameters for both stars, respectively: \teff = 5802, 5178 K and \lg = 4.33, 4.56.  They did not take into NLTE effects or hyperfine splitting in their spectral lines.  

\citet{Neuforge_1997} performed a differential analysis relative to the Sun
for $\alpha$ Cen A and B.  While they used ATLAS9 atmospheres \citep{Kurucz:2005p4698}, they used their own code for the electron pressures, gas pressures, opacities, and surface gravities.  They determined the stellar parameters for both stars as \teff = 5830, 5255 K and \lg = 4.34, 4.51, respectively. When analyzing the abundance ratio measurements for the elements by \citet{Neuforge_1997}, we found that their determinations were inherently different than those presented in the other works discussed here.  Namely, their abundances was consistently outside the range of values measured by the other literature sources by an average of 0.04 dex (later defined as the {\it spread}, see \S 3) for six elements. We attribute this dramatic difference to the author's use of their own code within their stellar models and/or the admitted problems with the weather instruments during the time of observations. We have therefore opted not in include this dataset within our analysis.

Finally, \citet{Laird:1985p1923} determined the carbon and nitrogen abundances with intermediate resolution ($\Delta \lambda$ = 1 \AA).  Surface gravities were calculated via the spectra and Str\"omgren photometry, augmented by gravities based on parallax data and estimated masses.  A differential analysis was performed and standard LTE via MOOG \citep{Sneden:1973p6104}.  The stellar parameters for each stars, respectively, were found to be \teff = 5600, 5030 K and \lg = 4.20, 4.43.  An analysis of their abundance ratios found that [C/Fe] was consistently offset by 0.2 dex and [N/Fe] by -0.65 dex, as a result of their stellar atmospheres being too cool.  We found that this analysis was not consistent with the other five catalogs and have therefore chosen not to include the abundances here.

Our analysis has yielded five literature sources: \citet{AllendePrieto:2004p476, Gilli:2006p2191, PortodeMello_2008,Thevenin:1998p1499, Valenti:2005p1491}, with similar analyses (for example, predominantly curve-of-growth and all using LTE, as opposed to NLTE corrections), stellar atmospheres, and data corrections (hyperfine structure was largely ignored).  We have compared these datasets with respect to one another in order to rule out any systematic offsets that may be present.  We found that to a reasonable degree, the datasets were comparable, with the exclusion of \citet{Neuforge_1997} and \citet{Laird:1985p1923} as previously mentioned.  We also investigated \citet{Valenti:2005p1491} in particular, since their method of analysis involved SME as opposed to curve-of-growth.  Despite previous claims that SME produces results that vary from other methodologies, we found that for $\alpha$ Cen A and B, this was not the case.

\section{Stellar Abundances in A \& B}
Using the element abundance ratios from the five catalogs, we are able to analyze 25 elements plus iron in both $\alpha$ Cen A and B, see Fig. 1 (left).  In an attempt to make the datasets more copacetic, we also have placed the abundance ratio measurements on the same solar scale. As an example, \citet{Gilli:2006p2191} determined that the abundance ratio for [Ti/H] = 0.28 dex for $\alpha$ Cen A using the solar scale by \citet{Anders:1989p3165}, where $\log \epsilon(Ti)$ = 4.99.  We wish to renormalize using the solar abundances of \citet{Lodders:2009p3091}, where $\log \epsilon(Ti)$ = 4.93.  Therefore, the renormalized value of [Ti/H] $= 0.28 + 4.99 - 4.93 = 0.34$ dex.  This renormalization allows us the only correction available that did not require the recalculation of the individual abundances.
In the instance where multiple catalogs measured the same element within one of the stars, we have chosen to use the median value.  In this way, we do not favor any one catalog and also avoid the presence of outliers and systematic offsets.  

We do not wish to gloss over the abundance ratio variations between catalogs, the largest of which we call the {\it spread}, or the maximum determination minus the minimum.  We have therefore plotted the abundance ratios in Fig. 1 (left) with error bars that are indicative of the spread in the data between catalogs in order to determine the upper bound in uncertainty.  For those cases where only one catalog measured the star for a particular element, we used the respective error, see Table 1.  

The most apparent result from Fig. 1 (left) is the similarity between the abundance ratios within the binary stars, as well as solar (dotted line).  The average abundance ratio for all the elements measured within $\alpha$ Cen A is 0.002 dex, while the elements within $\alpha$ Cen B have a mean of 0.03 dex.  In other words, both have element abundance ratios that are generally solar, with the B-star abundances slightly higher on average than the A-star.  

Analyzing the relative abundances within the two stars with respect to each other, we found that the average of the absolute difference, or $\mid B_ {abunds} -$ A$_{abunds}\mid$, is 0.05~dex with a formal 1$\sigma$ uncertainty of 0.05~dex.  We use the absolute difference in order to correctly account for both positive and negative differences between the stars.  The mean for the abundance ratio uncertainties (both respective error and spread) in $\alpha$ Cen A and B are 0.05 dex and 0.06 dex, respectively. As a further check of any statistically significant difference between the relative A/B abundance ratios, we performed a $\chi^2$ test. The test resulted in a $\chi^2$ of 32.6 for 26 degrees of freedom which is equivalent to a 17\% chance that the observed results diverge from each other by chance. This is not a statistically significant result and thus we conclude that the abundance ratios for the two stars are generally similar to both each other and to solar, with the average difference on the order of the average error. The abundances ratio within the two stars do vary on a case-by-case basis, as shown in Fig. 1 (left).  A total of 17 out of the 26 elemental abundance ratios have a difference greater than the average difference (0.029 dex).  For 6 of these elements, the difference is greater than the associated uncertainties: Al, Ca II, Ti II, V, Y, and Eu.

\begin{table*}\scriptsize 
\caption{Abundances [X/Fe] and Uncertainties in $\alpha$ Cen A \& B}
\begin{center}
\begin{tabular}{l|ccccccccccccc}
\hline
\hline
Elements & C &  O &  Na &  Mg &  Al &  Si &  Ca &  CaII &  Sc &  ScII &  Ti &  TiII &  V \\
\hline 
$\alpha$ Cen A & -0.06 & -0.01 & 0.06 & 0.09 & -0.01 & 0.01 & 0.02 & 0.15 & -0.01 & 0.12 & -0.03 & 0.10 & -0.06 \\ 
Uncertainty (A) & 0.04$^{\dagger}$ & 0.07 & 0.01 & 0.08 & 0.03$^{\dagger}$ & 0.05 & 0.06 & 0.01$^{\dagger}$ & 0.05 & 0.01 & 0.07 & 0.03$^{\dagger}$ & 0.09 \\
$\alpha$ Cen B & 0.01 & 0.02 & 0.08 & 0.09 & 0.06 & -0.06 & -0.10 & 0.19 & 0.03 & 0.13 & -0.02 & 0.21 & 0.11 \\ 
Uncertainty (B) & 0.04$^{\dagger}$ & 0.04$^{\dagger}$ & 0.06 & 0.05 & 0.03$^{\dagger}$ & 0.10 & 0.16 & 0.01$^{\dagger}$ & 0.03$^{\dagger}$ & 0.15 & 0.07 & 0.03 & 0.03 \\
\hline Elements &  Cr &  Mn &  Fe$^{*}$ &  Co &  Ni &  Cu &  Zn &  Y &  YII &  BaII &  Ce &  Nd &  Eu \\
\hline 
$\alpha$ Cen A & -0.05 & -0.05 & 0.28 & 0.03 & 0.01 & -0.10 & 0.21 & -0.06 & -0.12 & -0.16 & -0.01 & -0.10 & -0.20 \\ 
Uncertainty (A) & 0.01 & 0.01 & 0.16 & 0.05 & 0.07 & 0.08 & 0.04$^{\dagger}$ & 0.05 & 0.04$^{\dagger}$ & 0.01 & 0.11 & 0.10 & 0.06$^{\dagger}$ \\
$\alpha$ Cen B & -0.09 & -0.03 & 0.31 & 0.11 & 0.01 & -0.02 & 0.25 & 0.05 & -0.14 & -0.17 & -0.05 & -0.14 & -0.04 \\ 
Uncertainty (B) & 0.07 & 0.09 & 0.11 & 0.10 & 0.06 & 0.02 & 0.04$^{\dagger}$ & 0.04$^{\dagger}$ & 0.04$^{\dagger}$ & 0.05 & 0.04$^{\dagger}$ & 0.04$^{\dagger}$ & 0.06$^{\dagger}$ \\
\hline
\hline
\end{tabular}
\label{tab.abunds}
\end{center}
\raggedright{
{$^{*}$ Defined as [Fe/H].\\
$^{\dagger}$ Uncertainty as a result of respective error, as opposed to the spread in the data.}}
\end{table*}

\section{Nearby Abundance Implications}
The concept that chemical history could be understood via stellar compositions and dynamics first came from \cite{eggen_1962_aa}.  They determined that different metallicities corresponded to different parts of the Milky Way, such that: metal poor stars are within the halo, slightly less metal poor stars are within the thick disk, while the Sun and nearby stars are more enriched, more  ``average," in the thin disk.

Part of the allure of studying the $\alpha$ Cen system is due to the fact that it is the closest system to the Sun,
Taking into account that both the A and B components are solar-like in mass: 1.105 $M_{\odot}$ and 0.934 $M_{\odot}$ \citep{Pourbaix_2002}, we can assume a similar evolution as the Sun.  However, there is a distinctly noticeable difference in the chemical compositions of the $\alpha$ Cen system and the Sun with respect to the typical metallicity indicator.  For $\alpha$ Cen A and B, respectively, [Fe/H] = 0.28, 0.31 dex (see Table 1).
   
Given that $\alpha$ Cen A and B are binary stars, where one is a confirmed planet host and the other is not, we would expect to observe a signature of planetary formation on the abundances within an exoplanet host star.  The $\alpha$ Cen system proves an excellent case study for characterizing the abundances within hosts versus non-hosts.  However, we and the majority of authors who have analyzed the abundances in the $\alpha$ Cen system (see \S 1), have found that the abundances are rather variable.  Therefore, we regard the uncertainties on the abundance ratios within Table 1 as an upper bound.  We now analyze these differences with respect to the dynamic evolution of $\alpha$ Cen, as well as exoplanet host metallicities.

\subsection{Binary Formation Scenario}
Given the similarity in the stellar abundance 
ratios, yet super-solar [Fe/H],
we briefly discuss the
formation and dynamical evolution scenarios for the $\alpha$ Cen
system.  In relation to the Sun, the $\alpha$ Cen components are
slightly older \citep{Mamajek_2008} and have comparable heliocentric
space velocity components relative to the solar neighborhood
\citep{Holmberg_2007}. The similar abundance ratios of the $\alpha$ Cen
components indicate though
that they are typical thin-disk stars
which may have formed from the same
material as the Sun \citep{Freeman_2002}. The question then arises as
to whether the components themselves formed together or separately.

The current understanding of binary formation mechanisms favors a mutual formation process rather than a capture scenario since the latter requires a conservation of energy that is difficult to achieve without the involvement of a third body \citep{Boss_1992}. A complete Keplarian orbital solution for the A and B components is provided by \citet{Pourbaix_2002} which contains an eccentricity of $e = 0.5179 \pm 0.00076$. Although this is a high eccentricity for the system, it is not atypical for binary systems with long periods. In fact, \citet{Duquennoy_1991} show that this falls near the peak of the eccentricity distribution for binaries with periods larger than 1000 days. The assumption that the $\alpha$ Cen system formed from the same material is consistent with the similar relative abundance ratios of the components. It is therefore unlikely that the $\alpha$ Cen system underwent a capture scenario for the two primary components as this requires significant multi-body interactions early in it's history.

\subsection{Exoplanet Host Metallicity}
The planet-metallicity correlation was first put forward by
 \citet{Gonzalez:1997p3950} and was later refined by \citet{Fischer:2005p948} who 

 found that the probability of gas giant formation went as the square of the number of metal (or [Fe/H]) atoms.  Juxtaposed to \citet{Fischer:2005p948}, \citet{Buchave_2012} noted that the metallicity range of stars hosting terrestrial ($R_p < 4.0R_{\oplus}$) exoplanets is relatively large $-0.6 < [m/H] < +0.5$, where [m/H] is the amount of non-hydrogen and -helium abundances within the stellar atmosphere.  
This metallicity range corresponds to the [Fe/H] range observed in the thin-disk stars, implying that presence of terrestrial exoplanets may be extensive in the local neighborhood.
However, \citet{Buchave_2012} also argued that the average metallicity is lower for stars hosting terrestrial planets than stars hosting gas giants.  With respect to the $\alpha$ Cen system, we find that the [Fe/H] abundance in the $\alpha$ Cen components are comparable.

One of the pitfalls of analyzing the [Fe/H] content or more generic [m/H] is detail lost in the generalization, making it difficult to determine the underlying connection between stellar metallicity and the presence of exoplanets.  Solar twins that host terrestrial planets reflect a relative deficiency in the refractory elements with respect to the volatile elements on the order of $\sim20\%$ or $\sim$0.08 dex in comparison to the Sun \citep{Melendez:2009p7788,Ramirez:2009p1792,Ramirez_2010}.  This deviation is possibly linked to presence of terrestrial exoplanets, where refractory elements (with condensation temperatures, T$_c$ $\gtaprx 900 K$) within the solar convective envelope were preferentially accreted onto protoplanetary dust grains and therefore depleted in the host star.  However, both $\alpha$ Cen A and B are more enriched in [Fe/H] than solar (Table 1).  Following the discussion in \citet{Ramirez_2010} regarding HD 160691 and HD 1461, we note that the difference in chemical evolution changes the interpretation of abundance ratios, especially with regard to the volatile elements.

We have plotted the abundance ratios for both $\alpha$ Cen A and B from Table 1 with respect to T$_c$ as given in \citet{Ramirez_2010} in Fig. 1 (right).  The abundance ratio trends of [X/Fe] vs. T$_c$ are more robust for T$_c$ $\gtaprx 900 K$, where the trend becomes non-linear below T$_c \sim$ 1000 K.   The refractory elements within $\alpha$ Cen A have a mean of -0.02 dex and a 1$\sigma$ uncertainty of 0.09 dex.  Similarly, within $\alpha$ Cen B the mean is 0.01 dex with a 1$\sigma$  uncertainty of 0.11 dex.  We have also plotted a linear fit (solid lines) for the refractory elements only (T$_c$ $\gtaprx 900 K$), disregarding the abundances of C, O, S, and Zn, for $\alpha$ Cen A and B in Fig. 1 (right).  These fits give a slope of 0.00015 $ \times 10^{-3}$ dex K$^{-1}$ for $\alpha$ Cen A and -0.0094 $ \times 10^{-3}$ dex K$^{-1}$ $\alpha$ Cen B.  

We find that these slopes align well with the analysis in \citet{Ramirez_2010}, for example their Fig. 9 and with regard to HD 160691 and HD 1461, where they noted that iron-rich stars show slopes near zero or below.  Per their interpretation for planet formation indicators within the abundance ratios, this confirms the signature of a terrestrial planet orbiting $\alpha$ Cen B and implies that $\alpha$ Cen A hosts a terrestrial planet yet to be discovered.   The lack of a confirmed exoplanet orbiting $\alpha$ Cen A, which may be due to detection limitations, makes any conclusion regarding planet formation in this system preliminary at best.  Due to the similar relative abundances observed in $\alpha$ Cen A as compared to B, it seems unlikely that a planet may have been accreted onto $\alpha$ Cen A.

\section {Conclusion}

The combined abundance measurements from \citet{AllendePrieto:2004p476, Gilli:2006p2191, PortodeMello_2008,Thevenin:1998p1499, Valenti:2005p1491} allowed us to better analyze the chemical evolution and formation history of both $\alpha$ Centauri A and B.  We found that abundance ratios within both of the stars were in general solar, where the mean was 0.002 dex and 0.03 dex, respectively, regardless of super-solar [Fe/H] measurements.  More physically this suggests that the  $\alpha$ Centauri system was formed from similar material as the Sun.  In addition, the average of the absolute difference between the two stars was 0.05 dex, such that $\alpha$ Cen B is slightly more enriched than $\alpha$ Cen A.

The abundance ratio determinations for $\alpha$ Cen A and B imply that both components were formed during the same epoch from the same or similar protostellar cloud rather than a capture scenario. The age and spatial velocity is comparable with solar, although 
the $\alpha$ Cen system and the Sun are also unlikely to have formed together. 
The Keplerian orbital parameters of the system is fairly typical of binary systems with relatively large orbital periods and is in a stable configuration on the timescale of planet formation scenarios.

If $\alpha$ Cen A and B were formed from the same material, this proves an excellent place to study the effects of hosting a terrestrial exoplanet on the stellar abundances.  There was little statistical deviation in the abundance ratios between the two stars, where terrestrial planetary formation theories predict some offset.  We found that the refractory abundance ratio measurements in $\alpha$ Cen B are relatively solar and similar to those observed in $\alpha$ Cen A, both with linear fit slopes near-zero or negative.  Our results, combined with the analysis of \citet{Ramirez_2010}, confirm that the abundance ratios in $\alpha$ Cen B show the signature of the confirmed planet and suggest that $\alpha$ Cen A is likely a terrestrial planet host.

In order to better examine the correlation between the metallicity found within giant exoplanet hosts and non-hosts, there has been a slew of recent surveys, i.e. \cite{Bond:2008p2099,Bond:2006p2098,Fischer:2005p948,GalvezOrtiz:2011p6730,Gonzalez:1997p3950,Laws:2003p3002,Reid:2002p4078,Santos:2004p2996,Santos:2001p6866,Sousa:2011p6355}.  The independent conclusions of these analyses is that stars with orbiting giant exoplanets are more iron-rich than non-host stars, however the results for the other elements are more discrepant between authors.  And unfortunately, the relatively small sample size for nearby terrestrial planets makes any sort of abundance characterization tentative.   The results of \citet{Buchave_2012} shows the abundance delineation between terrestrial hosts and non-hosts are far more subtle than for giant planet hosts.  Recognizing that the key to understanding planetary formation lies within stellar archeology, we look towards studies and compilations that are able to measure the individual element ratios within nearby stars.  It is through this level of detail that we will better understand the chemical evolution of our solar neighborhood.

\section{Acknowledgments}
The authors would like to thank David Ciardi, Nairn Baliber, Patrick Young, and Klaus Fuhrmann.  Also, they acknowledge financial support from the NSF through grant AST-1109662.  NRH also thanks CHW3 for his insight.

\label{lastpage}

\clearpage

\end{document}